# Security & Privacy Product Inclusion

DAVE KLEIDERMACHER, Google, USA
EMMANUEL ARRIAGA, Google, USA
ERIC WANG, Google, USA
SEBASTIAN PORST, Google, Germany
ROGER PIQUERAS JOVER, Google, USA

## Abstract

In this paper, we explore the challenges of ensuring security and privacy for users from diverse demographic backgrounds. We propose a threat modeling approach to identify potential risks and countermeasures for product inclusion in security and privacy. We discuss various factors that can affect a user's ability to achieve a high level of security and privacy, including low-income demographics, poor connectivity, shared device usage, ML fairness, etc. We present results from a global security and privacy user experience survey and discuss the implications for product developers. Our work highlights the need for a more inclusive approach to security and privacy and provides a framework for researchers and practitioners to consider when designing products and services for a diverse range of users.

## 1 Introduction

### 1.1 Executive Summary

Product inclusion is growing in importance in the hearts and minds of technology developers, who aim to ensure their products and services work well for everyone. Certainly, this is a huge focus for Google and is aligned with its corporate mission. Accessibility requirements for the billion+ disabled people and the focus on equitable facial recognition and rendering performance are salient examples. However, an area that arguably has been underrepresented is the impact of product inclusion across the gamut of security and privacy capabilities and strength of function for users. This paper draws on the experience of the world's most popular operating system, Android, with billions of users across all demographic representations and the diverse product offerings across Google's connected products and services to survey challenges in this domain. We show how we apply threat modeling, similar to that applied in the security domain itself, to uncover novel product inclusion risks and countermeasures that we hope will not only improve the security and privacy protections for users of Google products but also help other product developers better consider product inclusion in their security and privacy goals. The results of our survey demonstrated that the ability of users to achieve a high level of privacy and security is sometimes hindered by changes in the specific environment in which they live their digital lives. For example, a user who lives in an area that lacks consistent and cost-efficient mobile network connectivity may not be able to install security updates reliably or as often, weakening their security posture. However, this is one noticeable headwind; we survey many more areas, and we hope this work will encourage other researchers and product security and privacy teams to bring product inclusion to the forefront of their design and priority space.

## 1.2 Threat Modeling for Inclusive Security and Privacy

In security threat modeling, we aim to think expansively of the resources under protection and how adversaries may attempt to circumvent that protection to gain unauthorized access to or otherwise compromise those resources. Similarly, for product inclusion, we ask the question: how can the mere operation within a user's specific environment (rather than a malicious threat actor) lead to the degradation or failure of those protections?

The authors embarked upon a typical expansive exploration of potential environmental threats and came up with the following (non-exhaustive) list:

- **Income disparity:** Low-income demographics are associated with lower-cost phones with less capable hardware, including hardware for security, such as biometric sensors, security chips, cryptographic accelerators, and sensor mute switches. The degraded hardware can, in turn, lead to poor security outcomes.
- **Location disparity:** In specific geographies, poor connectivity may inhibit security updates (mentioned earlier) and encourage unsafe practices such as sideloading apps from untrustworthy sources.
- **Device sharing:** In specific demographics, it may be a social norm to share a single device amongst multiple family members; shared device usage may pose privacy risks in which one member of the family who requires privacy has their personal information exposed to another family member, either accidentally or intentionally.
- **ML fairness:** Algorithms may be trained on personal information (such as biometrics) from one demographic set that does not accurately reflect another demographic, leading to poor outcomes. A salient example is training a facial recognition algorithm used for device authentication on data collected from a majority demographic and thereby having that algorithm execute poorly with minority demographics, leading to higher false acceptance and/or rejection rates, leading not only to poorer usability (which may in turn cause some users to turn off device authentication completely, reducing security) but also making the biometric easier to spoof.
- **Security UI:** Consents, permission dialogs, and other security and privacy-specific user interfaces may be poorly worded and translated into all languages, leading users of those languages to make poor security and privacy decisions.
- **Application safety:** In some geographies, access to trusted app stores like Google Play may be lacking, leading users to download apps from less safe sources and putting their digital lives at risk. A recent example of this is Iran, where national [authorities have been reported recently to inhibit access to the Google Play store](#).
- **Age and fingerprints:** Is fingerprint fidelity reducing with age, making their use for biometrics less secure?
- **Mental acuity, disability, and security UI:** For users with less ability to manage complex user interactions, which security functions are more likely to be turned off or misused?  For example, if someone loses their phone, will they be able to navigate the complexity of a find-my-device function via a web interface or a family member's phone?
- **Risky reuse:** In some demographics, are users more likely to hand down or resell a phone to another user, extending the device's usage beyond its security support lifetime?
- **Repossession:** In some geographies with unusual business models, are devices more likely to be locked by the device retailer (e.g., due to a missed payment), causing loss of personal data for the user?
- **High risk users:** At certain times, users may be at significantly higher security and/or privacy risk than a typical user. Examples include politicians running for office, someone experiencing intimate partner violence, reporters, and dissidents. Given their outsized impact in these situations, what security and privacy measures must be considered?

The above exploration is by no means exhaustive. Over time, we hope that the product security community will develop effective methods for measuring relative security quality across device experiences, enabling users to make better decisions to protect their digital safety while providing a more well-lit path for device and software manufacturers to deliver more effective security and privacy solutions to the broadest possible range of demographic environments. As part of the work on this survey paper, we performed a global security and privacy user experience (SPUX) survey (Figueroa, 2022), asking a variety of questions related to the aforementioned threat model exploration, especially to explore whether specific geographical differences may correlate to significant differences in the ability of users to achieve higher levels of security and privacy. The same questions were asked of N=2000 consumers in the following nations: Brazil, Germany, India, Indonesia, Japan, Kenya, Nigeria, and the US. We will share salient results in applicable sections throughout the rest of this paper.

## 1.3 Shared Device Usage

As mentioned earlier, an inclusive privacy concern worthy of additional research and attention is reduced privacy caused by shared device usage that may be more common in specific demographics. The SPUX showed a strong indication of such disparity:

| Country | Economic Status | % of smartphone users that share a phone with someone in their household |
| --- | --- | --- |
| Japan | Developed | 8 |
| Germany | Developed | 9 |
| Brazil | Newly industrialized | 31 |
| India | Newly industrialized | **73** |
| Indonesia | Newly industrialized | 30 |
| Kenya | Developing | 40 |
| Nigeria | Developing | 39 |

While one might assume that sharing is more prevalent in lower-income nations, these results show there may be a cultural element that cannot be explained by income alone. At the same time, developing countries (Nigeria and Kenya) have a higher rate of sharing than developed countries. In most newly industrialized countries, one newly industrialized country (India) has a sharing rate approximately double that of the developing countries.

[Matthews et al. published a taxonomy of common household sharing modalities, including "borrowing," where Android's guest mode is a valuable approach to limit privacy risk, and "mutual use,"](#) where multiple household members share a single device for primary, everyday usage. This research also touches on the privacy concerns related to sharing.

The cultural expectation of device sharing and related privacy concerns in Southeast Asia nations India, Pakistan, and Bangladesh are reviewed by [Sambasivan et al.](#), with a particular emphasis on women and the influence of cultural sharing expectations on personal privacy. The research references other pertinent studies and reports on techniques women use to manage privacy amid device sharing.

Smartphones were developed as single-user devices, where a device lock (e.g., pin, password, and/or biometric) is used to protect against unauthorized physical access. Therefore, if a single phone is used communally, the device lock must either be turned off or shared with communal users. In either case, privacy will be reduced unless some other mechanism can further compartmentalize access between communal users.

Android supports [multiple users](#), which provides robust security and privacy boundaries between multiple users on a single smartphone, but not all manufacturers enable this (the feature is optional). The multi-user implementation requires significant additional RAM utilization, which may render the feature impractical on [entry-level devices](#). Optimizing multi-user implementations for system health and overall user experience is an active area of research and development for Google's Android team. We intend to make this privacy feature more inclusive for entry-level devices over time, addressing the sharing needs evident in the preceding table.

App developers have created alternative solutions with varying levels of security robustness for multi-user compartmentalization. Some app lockers add a separate PIN to guard access to selected apps. Another popular approach is creating a separate folder with secondary instances of commonly used apps (e.g., Whatsapp messenger) with secondary user accounts. This secondary folder can be assigned a separate access control, such as a PIN, to provide privacy between communal users. Unfortunately, these app "containers" often share storage and permissions across all the apps stored within the container, violating Android's standard app isolation properties. The authors encourage further research to assess the security and privacy strength and usability of the various intra-device shared-use compartmentalization technologies, including multi-user, account profiles, app lockers, and container solutions.

## 1.4 Physical Theft

One of the best-known risks to smartphone users is device theft, and the device lock with associated data encryption has long been the most critical countermeasure to protect the device users' data from theft. As part of SPUX, we looked at whether device theft was more prevalent in specific geographies and whether a device lock met the increased risk (hopefully). In the early years of smartphones, device locking lacked the convenience of biometrics, especially fingerprint readers, leading many users to leave their devices unlocked rather than endure the relative inconvenience of knowledge factors such as passwords and PINs.

As seen in the table below, users in the developing countries of Africa surveyed were much more likely to have been affected by device theft than developed countries. However, the good news is that biometric authentication usage was not significantly lower in the developing countries:

| Country | Economic Status | % of smartphone users that have had a device stolen at some point | % of smartphone users that often/always use biometrics for device lock |
| --- | --- | --- | --- |
| Japan | Developed | 2 | 88 |
| Germany | Developed | 9 | 87 |
| Kenya | Developing | **58** | 89 |
| Nigeria | Developing | **56** | 91 |

This result is one of the most gratifying of SPUX success for the Google Android team, which has been working closely with the Android OEM ecosystem for many years to increase device lock adoption, with a specific focus on making biometric sensors more commonly available across the range of device price points while also improving the overall biometric user experience within the operating system.

Another important countermeasure for stolen devices is the ability to remotely locate, lock, and wipe stolen devices. Google's [FindMyDevice](#) service is an example of technology users may employ for this protection. However, more research is needed to assess whether users are sufficiently aware of these capabilities and how to use them. In addition, Google and other smartphone technology developers are actively working on improving privacy and user experience in this area.

## 1.5 Accessibility Services

Android is constantly innovating in [accessibility services](#) that help people with disabilities make the best use of their mobile devices. Unfortunately, malware authors also leverage accessibility services to abuse users. Socially engineering users into enabling accessibility services is the second most impactful abuse vector employed by malware authors, as shown in the table below, which reflects a 2022 internal Google analysis of top malware abuse vectors:

| Top Malware Abuse Vectors, by Priority (highest to lowest) |
| --- |
| Dynamic Code Loading |
| Observing and interacting with apps using an accessibility service |
| Obfuscation |
| Data collection (SMS, contacts, location, photos) |
| Obtaining incoming SMS via notification listener |
| Obtaining login credentials & cookies from WebView |
| Obtaining login credentials via fake UIs (in apps) |
| Cloaking based on device/geo/carrier/date |
| Obtaining incoming SMS via READ_SMS permission |
| Background UI launches |
| Play apps prompting user to install off-Play apps |
| Apps distributed via direct web URL downloads |
| Use of low target API level |
| Redirecting outgoing calls |

Android and the Google Play store have numerous mitigations to protect users from improper use of accessibility services. However, the risk of abuse from malware increases when users already have accessibility services enabled and do not need to fall prey to an attacker's attempt to convince users to enable them. While we expect some variation in the adoption of accessibility services across various demographics, we were surprised by the large delta in adoption across geographies in the SPUX survey, notably:

| Country | Economic Status | % of smartphone users that make use of accessibility features |
|---|---|---|
| Japan | Developed | 9 |
| Germany | Developed | 7 |
| Kenya | Developing | **78** |
| Nigeria | Developing | **77** |

The most commonly used accessibility features in the SPUX survey were adjusting text size, voice-activated input, and interacting with Android's screen reader. These results show the importance of investing in counter-abuse techniques and technologies that address the accessibility services attack vector (which will remain a priority for Google's Android team).

## 1.6 Price Points and Hardware Security

In the preceding section, we related how adopting biometric authentication has succeeded dramatically. Still, one that took years to develop was made possible by the competition and reduced the price points of biometric sensors over time. However, security and privacy are impacted by a wide range of other hardware components. Therefore, they may also be exposed to product inclusion challenges where lower-priced devices can offer different hardware security and privacy capabilities than flagship devices. While a "minimum bar" of security applies to all Google-certified Android devices, there will always be a continuum of hardware capabilities across different price points that remain optional until hardware availability and other software compatibility constraints are solved at scale; this phenomenon of capability also provides opportunity for OEM differentiation and is similar across many feature areas, including display quality, memory size, battery life, and security.

The following table shows some important contemporary hardware-specific security features that are optional for Android OEMs and only available on some Android implementations and, therefore, may pose inclusive security challenges at different price points:

| Feature | Initial Android version availability | Mandatory for Android smartphones | Device availability/Notes |
| --- | --- | --- | --- |
| Memory Tagging Extensions (MTE) | Android 12 | future | Available in Google Pixel 8 as a developer option and other devices expected in 2024+ |
| File-based encryption | Android 7 | Android 10 | all |
| Verified boot | Android 4.4 | Android 7 | all |
| Multiple biometrics | Android 10 | N/A | Was possible prior to Android 10, but BiometricManager API made it much easier for developers to leverage. Examples include Google Pixel 7 and Samsung Galaxy Fold 4, which support fingerprint and facial recognition biometrics. No plan to mandate the availability of multiple biometrics on a single device. |
| CHERI | future | future | |
| Android Virtualization Framework | Android 13 | Android 15 (planned) | Available in Google Pixel 7 and higher, and other devices expected in 2024+ |
| StrongBox keys | Android 9 | Android 14 | Required for Google Wallet compatibility |
| Remote Key Provisioning | Android 12 | Android 13 | |

Consider the role of multiple biometrics as an example of the enhanced security made possible by these optional features. The availability of multiple biometrics can provide more flexible options for any user. However, it may be beneficial for specific populations. For example, research has demonstrated significant differences in fingerprint-matching accuracy in elderly populations, who may also suffer from usability challenges (failure to capture or failure to enroll) that may degrade overall security. Suppose a device supports multiple biometric authentication methods for device lock. In that case, someone experiencing challenges with one form of biometric may switch to another to improve security strength. Furthermore, biometric authentication relies on heuristic matching algorithms that have been demonstrated to be susceptible to advanced spoofing attacks (such as creating a 3D mask from a photo that can be used to fool a facial biometric). Having multiple biometrics will provide users with an alternate authentication choice, an "insurance policy," in the event a significant weakness is discovered in one of the biometrics that cannot be feasibly addressed with a firmware update.

Another example of price-relevant security quality is using a secure element embodied by the StrongBox keystore type in the preceding table. Secure elements provide higher physical security for cryptographic keys and processing. Secure elements are typically certified against more stringent security standards than anything else on the typical smartphone SoC (e.g., one-time programmable memory). StrongBox-compatible secure elements may be added to a smartphone as a discrete component or included within the device's primary SoC. In either case, there is an extra cost associated with the technology, which is why the StrongBox feature has been optional and has, over time, become more widely available across price points. One example of SoC incorporating a certified secure element is Qualcomm's Secure Processing Unit (SPU) enabled SoCs, which can be found in higher-end offerings such as the Snapdragon 7 and 8 series but not the Snapdragon 6 series.

In addition to security-specific hardware, numerous examples of security inclusion differences are caused by non-security-specific hardware capabilities. For example, a device with less flash memory may be more likely to fail a security update due to a lack of storage to temporarily store the update while it is being applied. Another example is the availability of NFC in contactless payments. While NFC-based contactless payment has been available in flagship smartphones for many years, some low-cost devices still need NFC hardware for contactless payments. A smartphone user unable to make contactless payments must carry cash or credit cards, increasing their risk of physical theft. In 2023, Google decided to address the gap in NFC contactless capabilities in many smartphones in Brazil by adding support for QR code-[based payments](), thereby addressing this personal risk inclusion gap.

In addition to the cost of optional hardware components (e.g., biometric sensors, secure elements, and NFC chips), the variations in availability shown in the previous table are also caused by product family lifecycles and software constraints. One example is the adoption of file-based encryption, whose impact on smartphone performance is considerable when the SoC's native instruction set does not support bulk encryption offload (e.g., AES in hardware). For many years, FBE was commonly adopted on almost all devices, with the exception of the most performance-constrained phones, which still lacked the AES offload. In 2019, Android engineers developed a cryptographic construction called Adiantum, which performs better than AES when it cannot be hardware offloaded, precluding the need for the carveout (finally).

Nevertheless, cost remains an important consideration that can affect the ability of users to obtain the best possible security capabilities. SPUX examined differences in typical smartphone price points across geographies. As shown in the table below, there is a significant disparity in the percentage of devices purchased above an approximate USD 300 price point, and therefore, access to the most advanced security features would also be impacted by this disparity.

| Country | Economic Status | % of smartphone users with devices below approx USD $300 (among those that identify their phone's price range) |
|---|---|---|
| Japan | Developed | 17 |
| US | Developed | 24 |
| Kenya | Developing | **79** |
| Nigeria | Developing | **62** |

This data may not be surprising, but it reminds the community of the importance of considering hardware-dependent security capabilities. One way Google is working to improve the situation, other than innovating in software security and infrastructure together with our partners, is to bring more transparency about the differences in optional hardware-dependent security features found across Android devices through independent security ratings or "scores" that will make it easier for users to make the best possible security decisions to fit their risk and cost requirements.

Another cost-related security concern is the use of metered data connections. Mobile devices depend on metered (i.e., cellular) mobile networks worldwide, but the cost and reliability of that connectivity vary. While the situation continues to improve over time (notably, India saw a [95% reduction](#) in mobile network consumer retail pricing in the five-year period ending in 2019, leading the nation to have the lowest mobile network pricing in the world), the concern about mobile data usage persists, especially in developing countries. In the past, the Android team has received anecdotal reports of consumers delaying or canceling security updates due to concerns about the use of mobile network data. SPUX asked consumers about their current level of concern, with the following salient results:

| Country | Economic Status | % of smartphone users "very concerned" about the use of mobile data for OS updates |
|---|---|---|
| Japan | Developed | 12 |
| Germany | Developed | 11 |
| Kenya | Developing | **60** |
| Nigeria | Developing | **59** |

This concern may lead consumers to leave their devices unpatched, increasing their security risk, yet another example of potential inclusion challenges in security and privacy.

# 2 ML Fairness and Product Inclusion

## 2.1 Introduction

Many Android security services, features, and core functions rely on ML to make decisions. ML drives analysis and automated decision-making in biometric authentication, abnormality/attack detection, and security process automation.

With that understanding, it's essential to ensure that when ML algorithms make these automated decisions, they do so in an inclusive way, resulting in equitable security outcomes for all users. Suppose the ML models have issues with fairness or built-in biases against underrepresented groups. In that case, the models may underperform (e.g., a higher error rate for people with a darker skin tone) or even make incorrect decisions concerning those users. This can lead to different forms of harm, from service quality issues to reinforcing or introducing stereotypes associated with underrepresented groups to ultimately reduced security protections and increased risk of being victimized in cyber security attacks.

## 2.2 General discussion on ML fairness

Examining these biases in ML models is familiar, and the ML and research communities are broadly exploring this area. Within Google's research, a few areas of bias have been identified (Google Developers, 2022). A crash course on these types of biases has been created by Google (Google YouTube, 2017). The following table summarizes some common ML bias types:

| Bias Type | Description |
| --- | --- |
| Reporting bias | Reporting bias occurs when the frequency of events, properties, and/or outcomes captured in a data set do not accurately reflect their real-world frequency. This bias can arise because people tend to focus on documenting circumstances that are unusual or especially memorable, assuming that the ordinary can "go without saying." |
| Automation bias | Automation bias is the tendency to favor results generated by automated systems over those that are generated by non-automated systems, irrespective of the error rates of the results |
| Selection bias | Selection bias occurs if a data set's examples are chosen in a way that is not reflective of their real-world distribution. Selection bias can take different forms such as:<br>● Coverage bias: When data is not selected in a representative fashion, this results in coverage bias.<br>● Non-response bias: Gaps in participation of the data-collection process can result in non-response or participation bias due to the data not being representative of the targeted population.<br>● Sampling bias: Absence of proper randomization techniques utilized in the collection of data can result in sampling bias. |
| Group attribution bias | The tendency of generalizing what is true of individuals to an entire group to which they belong is known as group attribution bias. This manifests in two key ways:<br>● In-group bias: Occurs when there is a preference for members of a group to which a party belongs, or for characteristics for which the party also shares.<br>● Out-group homogeneity bias: Occurs when there is a tendency to stereotype individual members of a group to which a party does not belong, or to see their characteristics as more uniform. |
| Implicit bias | When assumptions are made based on one's own mental models and personal experiences that do not necessarily apply more generally, this is implicit bias. |

## 2.2.1 Biases in model development

Bias can be introduced in each stage during the ML model development process. Common mistakes that can lead to bias are explained in (Google Developers, 2022) and subsequent literature on ML fairness (Radharapu, 2019). The phases of model development include defining the problem, collecting and preparing the data, training the model, evaluating the model, and integrating and monitoring the model. The diagram below breaks down the different biases that can be introduced at each process stage.

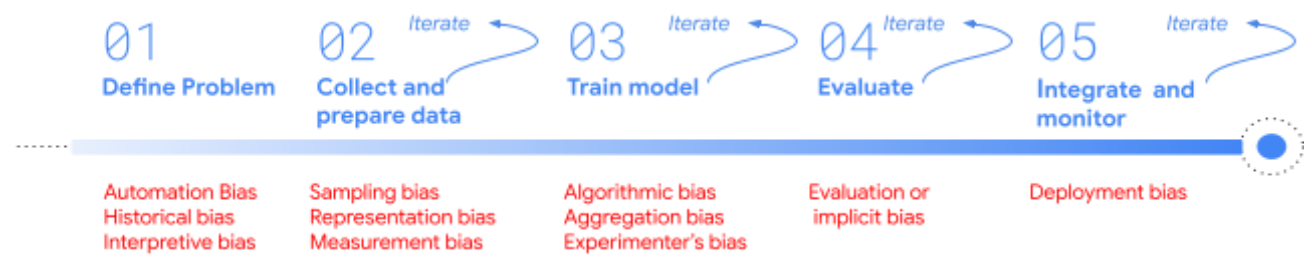

Figure 1: Bias In Each ML Development Stage (Radharapu, 2019)

ML model bias can cause the model to underperform for specific users and environments and lead to situations where wrong answers are given, or the model makes wrong decisions. A classification of these potential harms is shown below:

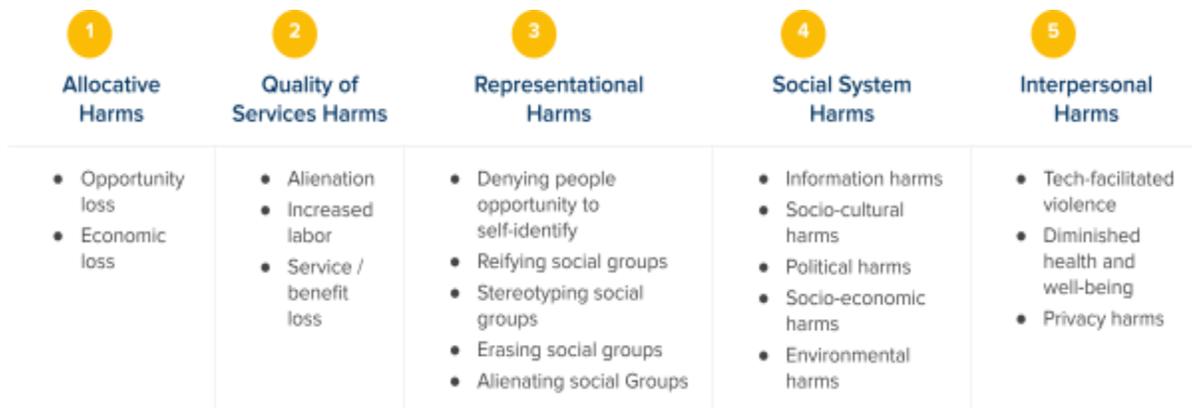

Figure 2: Types of Harms by ML Bias (Sheiby, 2019)

## 2.3 ML bias and impact on security and safety functions

As security and safety-related services increasingly use ML technologies, bias in ML can significantly impact users in terms of scale and severity. For example, according to (Godbole et al., 2022), American airports processed over 9 million international travelers in the first six months of 2021 using ML-based biometric systems. This means that even a tiny difference in the performance across groups may cause significant consequences for users.

A higher error rate for security and safety services can lead to service inconvenience and an increased risk of becoming a victim of cyber attacks. For example, model errors can lead to authentication bypasses and false negative detections for phishing and malware, which can subsequently lead to unauthorized disclosure of personal information, financial loss, or even personal safety harms if the digital system is used to control or influence physical systems, actions, or behaviors. Thus, ML fairness is critical for inclusive ML-based security and safety products and systems. The following are examples where bias in ML can directly impact security and safety.

### 2.3.1 Bias in biometric systems

Most biometric-based security and safety systems use ML technologies to some extent, such as face and fingerprint-based identity verification or authentication. Researchers have been concerned about fairness in biometric-based systems due to finding many bias-related issues over the past several years. For example, in the well-known Gender Shades" project (Buolamwini, 2018), it was found that three popular gender classification algorithms developed by several well-known vendors consistently performed worse on darker-skinned females, with error rates up to 34% higher than for lighter-skinned males (Figure 3). Additional examples can be found in (Bitzionis, 2019) and (Grother et al., 2019).

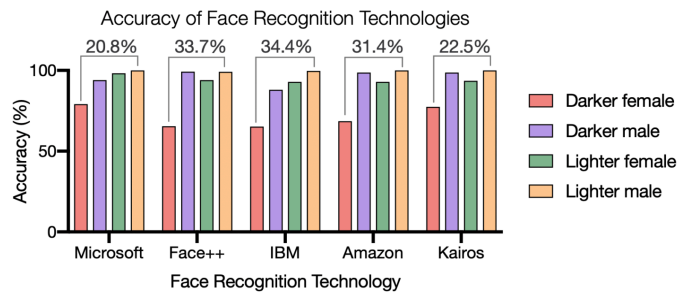

Figure 3: Racial Discrimination in Face Recognition Technology (Najibi, 2020)

Although significant progress has been made in the past years, many areas still need to be investigated thoroughly, and many issues still need to be addressed. (Drozdowski, 2020) provides a lengthy survey of research on fairness in biometrics systems. This survey, cited by multiple papers, identified key trends and challenges that still exist.

The major trends identified by this survey (Drozdowski, 2020) include:

- Most studies focused on fairness in face-based biometrics. There are significantly fewer studies on other modalities (e.g., fingerprint).
- Most studies concentrated on biometric recognition algorithms (primarily verification), followed by quality assessment and classification algorithms. However, some scenarios, such as anti-spoofing and attack detection, have yet to be investigated.
- The existing studies predominantly considered gender and race. The age covariate is considered but less frequently in the context of bias. Other covariates are often not addressed.
- Many studies focused on general accuracy rather than distinguishing between false-positive and false-negative errors. This is particularly important for security and safety-related systems because the two metrics have significantly different implications and are often used as critical benchmark metrics for security services.
- A significant number of studies conducted evaluations on sequestered databases and/or commercial systems. However, reproducing or analyzing their results may be impossible due to the unattainability of data and/or tested systems.

The survey also identified the following challenges in existing research:

- Existing research concentrates on specific statistical definitions, such as group fairness and error rate parity. Extending the existing estimation and mitigation works, such as considering other and more complex notions of fairness, could be considered crucial future work. Likewise, investigating tradeoffs between performance, fairness, user experience, social perceptions, monetary, and other aspects of the biometric systems might be of interest.
- Theoretical approaches must be pursued to demonstrate the bias-mitigating properties of the proposed methods.

- Isolating the effects of the demographic factors from other factors (i.e., environmental covariates like illumination and subject-specific covariates such as use of makeup) needs to be sufficiently addressed in many existing studies.
- More complex analyses could be conducted based on demographic attributes and combinations of different factors.
- Independent benchmarks for fairness measurement methodologies and metrics still need to be included. Security-specific metrics, such as False Reject Rate (FRR), False Match Rate (FMR) in face/fingerprint authentication, Attack Presentation Classification Error Rate (APCER), and Bona Fide Presentation Classification Error Rate (BPCER) in Presentation Attack Detection systems, should also be considered.
- Large-scale datasets designed specifically for bias-related research need to be collected.
- Humans are known to exhibit a broad range of biases. The influence of those factors on the biometric algorithm design, interactions with and use of biometric systems, and perceptions of biometric systems could be investigated.
- Social impacts are less studied, mainly for security systems, such as the ramifications of a false positive biometric match leading to a person being incarcerated.
- Some areas need more research, particularly fairness for Presentation Attack Detection (PAD) in face/fingerprint algorithms. PAD is a critical requirement for security/safety systems, but only a handful of papers have been published on this topic (Drozdowski, 2020).

## 2.3.2 Fairness in Face PAD (Presentation Attack Detection) Algorithms

Face-based authentication systems are widely used, from unlocking phones to identity verification in airports to controlling access to high-security buildings. A typical face authentication implementation consists of multiple subtasks, as shown in the diagram below:

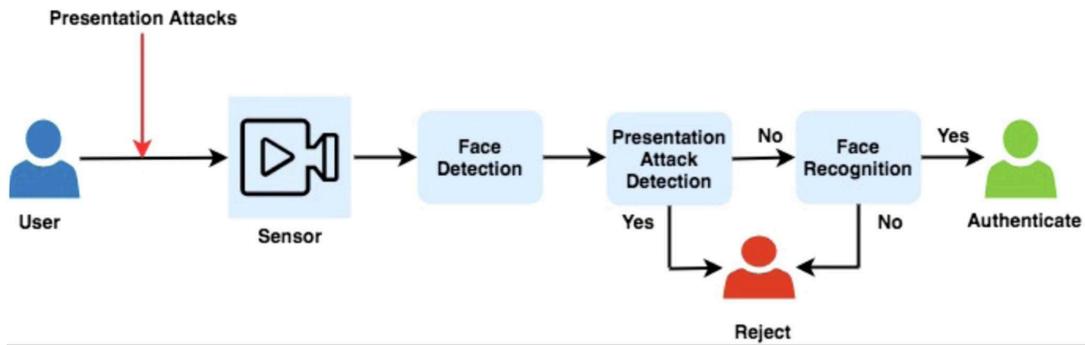

Figure 4: A typical face authentication flow (Abdullakutty, 2021)

In PAD attacks, an attacker attempts to use a printed photo, a replayed video, a 3D mask, etc., to impersonate someone or obfuscate their own identity. Given the widespread use of social media, it is often not difficult for an attacker to obtain someone's photo from the Internet. Therefore, if the face authentication system cannot detect a presentation attack, it can be easily bypassed, and its security performance will not be considered acceptable.

Although significant work has been done to address fairness in face recognition algorithms (Drozdowski, 2020), more must be done to address fairness with PAD. A recent fairness study on PAD was published in 2022 (Fang et al., 2022) in which the researcher found fundamental limitations from existing studies:

- Most studies focused on fairness issues in face recognition algorithms, with limited work on fairness for PAD.
- There is limited research on the fairness of PAD algorithms (the authors found only two papers on the topic)
- Existing research focuses on demographic covariates like gender and race. However, other factors, such as appearance traits, use of accessories like makeup, etc., are not considered.

A key reason for the limitations is that most publicly available PAD datasets do not contain information regarding demographic and non-demographic attributes, making it impossible to assess fairness. The researchers combined six PAD datasets out of the nine popular PAD databases to address those issues. They created a database (CAAD-PAD) with manually added non-demographic labels, including gender, beard (visible hair coverage around the mouth or shaved with only light hair roots), eyeglasses, bangs (hair covering more than 15% of the forehead), makeup (visible use of lipstick and eye shadow), long/short hair (hair beyond the shoulder), and curly/straight hair. Five algorithms, from texture-feature-based to deep-learning-based methods (LBPMLP, ResNet50, DeepPixBis, LMFD, LBP-MLP), are used in the study.

Performance of PAD is commonly measured by Attack Presentation Classification Error Rate (APCER), or the proportion of attack presentations incorrectly classified as bona fide presentations, and Bona Fide Presentation Classification Error Rate (BPCER), or the ratio of bona fide samples misclassified as attack samples. The researcher introduced a fairness metric called Accuracy Balanced Fairness (ABF), the weighted average of accuracy and fairness. Accuracy is defined as the proportion of correct predictions. Fairness is defined as the degree to which the model predicts equally well for all groups of people:

$$A(\tau) = \max(|APCER^{di}(\tau) - APCER^{dj}(\tau)|) / (1 - \max_D(APCER(\tau)), \forall di, dj \in D$$
$$B(\tau) = \max(|BPCER^{di}(\tau) - BPCER^{dj}(\tau)|) / (1 - \max_D(BPCER(\tau)), \forall di, dj \in D$$
$$ABF(\tau) = 1 - (\alpha A(\tau) + (1 - \alpha)B(\tau))$$

An ABF score of 1 means the evaluated model is accurate and fair. A score of 0 means that the model is neither accurate nor fair. A score between 0 and 1 means the model is somewhat accurate and fair. For example, an ABF score of 0.8 means the model is 80% accurate and 80% fair.

In the researcher's study, ABF is calculated under different decision thresholds $\tau$ = APCER, varying from 0.005 to 0.2. The study showed that most of the selected covariants will clearly influence the model and show a meaningful performance difference if the training does not consider fairness.

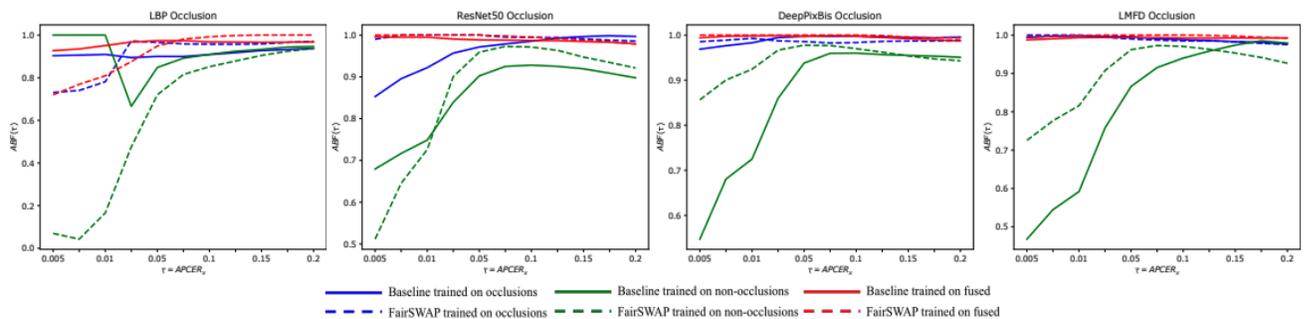

Figure 5: ABF values for PAD models on gender groups (Fang et al., 2022).

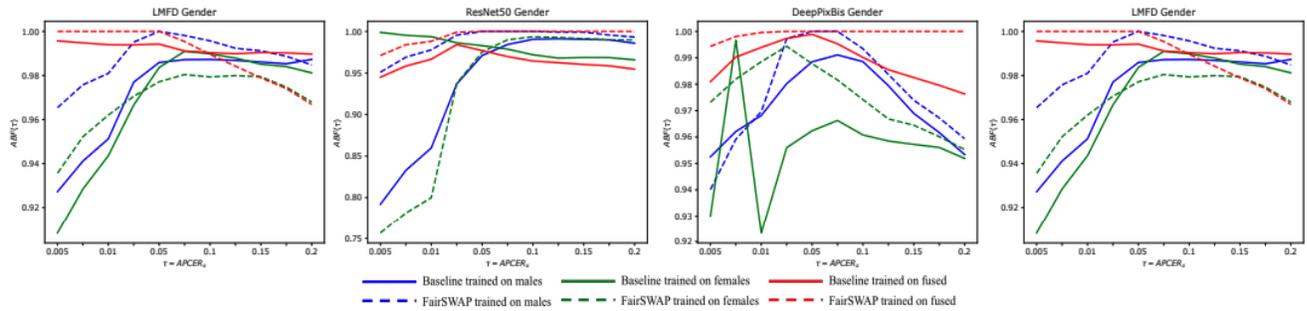

Figure 6: ABF values for PAD models on occlusion groups (Fang et al., 2022)

In Figures 5 and 6, higher and smoother lines indicate higher fairness. More results can be found in the paper (Fang et al., 2022). The study shows that models trained on separate gender groups possess more discrepancies than the same PAD models trained on fused data, and all models trained on fused data exhibit higher fairness. The researcher concluded that demographic and non-demographic factors can affect the fairness of face PAD. The training data and the ODTA (the threshold) can trigger bias in face PAD. The results also show that the performance change introduced by bias can be statistically significant. For example, when $APCER_x$ is between 0.5% and 1%, the BPCER for a PAD algorithm can have up to 6% difference in error rate for the gender test and up to 30% difference in error rate for the Bangs test. Since the BPCER directly influences the False Reject Rate (FRR), and the common FRR acceptable range for face recognition is less than 10%, the error rate added by bias may drive PAD performance beyond the unacceptable range.

The study did not directly evaluate the biases for APCER. APCER is essential because a higher APCER rate means an increased risk of impersonation by PAD attacks. Such attacks can impact users (for example, account authentication bypass) more than BPCER, which usually means service quality or DoS issues. Other limitations of the study include that the database is not designed for fairness study and has limited data for fairness evaluation; the attacks and the experiments are also relatively simple. However, as one of the few studies on fairness in face PAD, the research provided valuable insights into the gaps in existing research and issues that need further study.

### 2.3.3 Fairness for Fingerprint Systems

Compared to studies on fairness in face recognition systems, more research should be performed on fairness in fingerprint authentication systems. For example, a recent study (Godbole et al., 2022) published in 2022 only cited six papers in its section on related works. As the prevalence of fingerprint algorithms in authentication systems has increased, a fresh look at the fairness in state-of-the-art fingerprint systems has become urgent.

Looking across the current research, Marasco (Marasco, 2019) analyzed biases induced by sensors and image quality. Godbole et al. (Godbole et al., 2022) pointed out that bias in the demographic distribution of the dataset can produce biases, particularly for systems that use deep neural network (DNN) models. It also found a lack of a standard definition for fairness and a statistical framework to test it. For example, if a system's true match rate is 99.5% and 99.2% for males and females, respectively, how can we claim that the observed differential is statistically significant? How large should the dataset be if we want to claim with 95% confidence that the observed demographic differential is statistically significant?

The author proposed a statistical framework for fairness evaluation and performed evaluations on two private fingerprint databases.

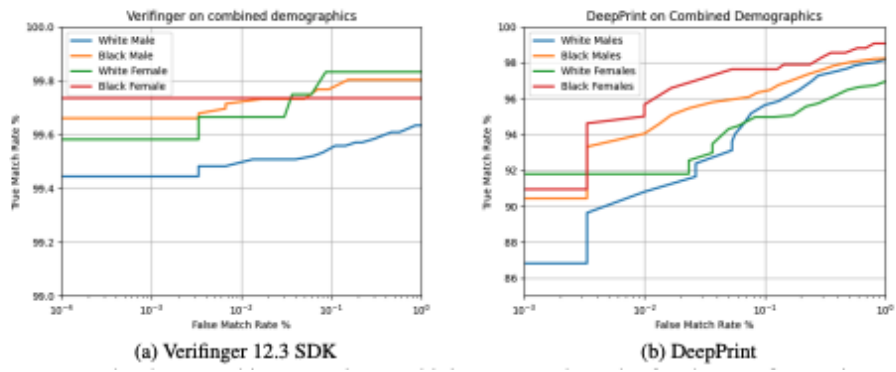

Figure 7: ROC Curves across the demographic categories combining race and gender for the two fingerprint matchers of database D1 (Godbole et al., 2022)

The results in Figure 7 show two trends:

1. More accurate algorithms are less likely to exhibit demographic-based bias
2. Most observed demographic differentials can be explained by the poor quality of some fingerprint images

In the study, out of the 9,112 total white subjects, if only 13 (0.15%) of the genuine similarity scores below the threshold were increased to be above the threshold, the bias between black and white subjects in Verifinger would disappear. The author claims that the study shows that no-demographic covariates are more significant in introducing bias. Still, the demographic covariates that cause bias in face recognition systems (particularly race and gender) may be less significant.

The concerns in (Google Youtube, 2017) are also reflected in (Marasco, 2019), a survey by Marasco on bias in fingerprint recognition systems in 2019. The research identified three types of bias:

| Bias Type | Description |
| --- | --- |
| Capture Bias | This includes types of sensors used to collect and sensors used to verify the images, environmental conditions (e.g., dry weather resulting in faint fingerprints), and the user interaction with the sensor. For example, the ergonomics of a fingerprint acquisition system (e.g., alignment, positioning and ease of use) affect the clarity of the ridge pattern. Such variations may lead to an increase of false rejection rate when at verification time a different fingerprint sensor is used. |
| Demographic Bias | Includes bias on gender and ethnicity. For example, one study found that 6.7% of very poor quality was associated with females while only 2.4% with males. Human skin is subject to a reduction of collagen when aging which may cause the skin to become wilting, drier and thinner, which can cause higher error rate (FRR) for elderly people. |
| Spoof Bias | The bias introduced by the materials (like silicone or gelatin) that attackers use to carry a print of a legitimate user and presented to a sensor for verification (Presentation Attacks or PAs). Existing PAD algorithms may not be trained by certain materials that attackers use and this can cause higher error rates. For instance, a detector trained to recognize a sheet of latex carrying the fingerprint of an individual may fail on a wood glue-based attack. In addition, existing studies only investigate cooperative spoofing methods which are not as realistic as the non-cooperative ones. For example, the algorithm is trained with data from cooperative users but is used to verify data from users who are not cooperative (attackers), or the spoof fabrication materials used for training and in actual attacks are different, or the sensor used for acquiring training data and to verify the data is different. |

The author concluded that a systematic approach is needed to address all types of bias, and a significant gap exists in current research. There are only minimal studies on spoofing bias, which were performed a long time ago without considering the latest attack methods and detection algorithms.

## 2.3.4 Fairness in Malware Detection Systems

ML plays a crucial role in malware detection. For example, in Google's Google Play Protect service for Android, ML detects a wide range of code and content-based abuses. ML will analyze each new app submitted to the Google Play app store, and a score is generated to indicate the probability that an app is abusive. If the score exceeds a certain threshold, the app is flagged for further action, which may include additional human review to confirm the abusive signal. Even if an app passes all checks at publish time, the ML systems will continuously monitor app behavior in case new signals or intelligence arise that cause the app to be flagged. The ML systems also analyze apps distributed outside the Google Play store, leveraging on-device processing, web crawlers, virus databases, and additional techniques to collect apps installed by sideloading, web URLs, alternative app stores, and other sources.

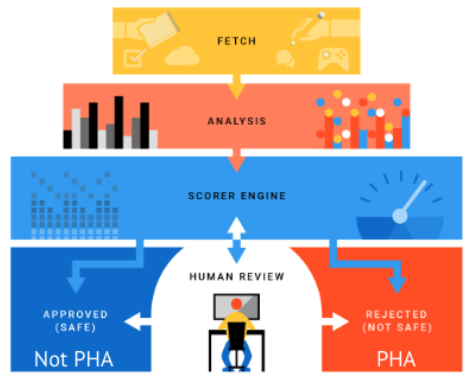

Figure 8: Google Play Protect ML Analysis Flow (Liderman, 2020)

Bias in ML-based malware detection has been studied for many years. For example, (Pendlebury et al., 2019) found that, based on an analysis of two popular classification algorithms on a large-scale dataset (AndroZoo), several popular Android malware classifiers are not evaluated in settings representative of real-world deployments. The study also identified two kinds of bias:

| Bias Type | Description |
|---|---|
| Spatial bias | Spatial bias refers to the unrealistic assumptions about the ratio of goodware to malware in the data. The ratio is domain-specific but should be enforced consistently in the training and testing phase to mimic a realistic scenario. |
| Temporal bias | Bias introduced by evaluations that include information from the future that wouldn't be available, or use test scenarios that don't reflect real-world conditions. In malware detection, this problem is exacerbated by families of closely related malware with the belief that including even one variant in the training set may allow the algorithm to identify many variants in the testing. |

The study also found that bias in training datasets can cause a statistically significant performance decrease with two well-known Android malware classifiers, DREBIN and MAMADROID (Figure 9).

Figure 9: F1-Score Impact of spatial and temporal experimental bias (Pendlebury et al., 2019)

The above findings are reflected in other research. For example, (Miranda et al., 2022) also suggest that the several popular datasets for Android malware scanning experiments are not representative of the population of Android applications in the real world. The researchers classified the bias in several popular datasets used for malware training into the following categories:

| Bias Type | Description |
|---|---|
| Spatial bias | The percentage of malware does not match the percentage of malware in the real world (e.g., too many malware apps and too few good apps in the dataset). |
| Class imbalance | The malware families in the dataset do not match the malware families in the real world. For example, some malware families have too many samples that include repacked malware samples within the same family, while other malware families have too few or no samples. |
| Sampling Bias | The good app population should match the real world good app population from a statistical perspective. Although it is impossible to collect all good apps in a dataset, the dataset should have enough diversified good apps to avoid false positives and classify good apps as malware. |
| Time bias | Malware evolves quickly. Adversaries not only create many new kinds of malware every year, they also use different obfuscating techniques to change signatures used by malware detection software (repacking). If a dataset is not updated frequently to reflect the active malware in the wild, spatial and class biases will be introduced even if it does not have such biases at the beginning. |

The researcher used six popular malware training datasets (Drebin, AMD, VS 2015, 2016, 2017, and 2018) for training and two groups of datasets randomly drawn from AndroZone (one for 2019 apps and one for 2020 apps) for testing. The researcher claims that the above biases can cause ML models to underperform under certain situations even if the models have good performance in general.

Some other challenges include:

1. Users in some regions tend to use alternative app stores more frequently. The relaxed security and verification practices in some stores can lead to a higher prevalence of malware. Malware families may also vary in different regions. Therefore, a dataset that is representative globally may have spatial bias or class imbalance bias for users in a specific region or specific group.
2. Malware may evolve at a different pace for different users because some users may use unpatched software or old phones with less secure designs. For example, in a survey on Android phone users (Figueroa, 2022), it is found that in the United States, some users (like teens) perform secure updates less often than others (figure 10), and some users (like older users) use old phones more than other users (figure 11). This may affect the malware evolution rates for such users, and a globally representative dataset may have time bias for those users.

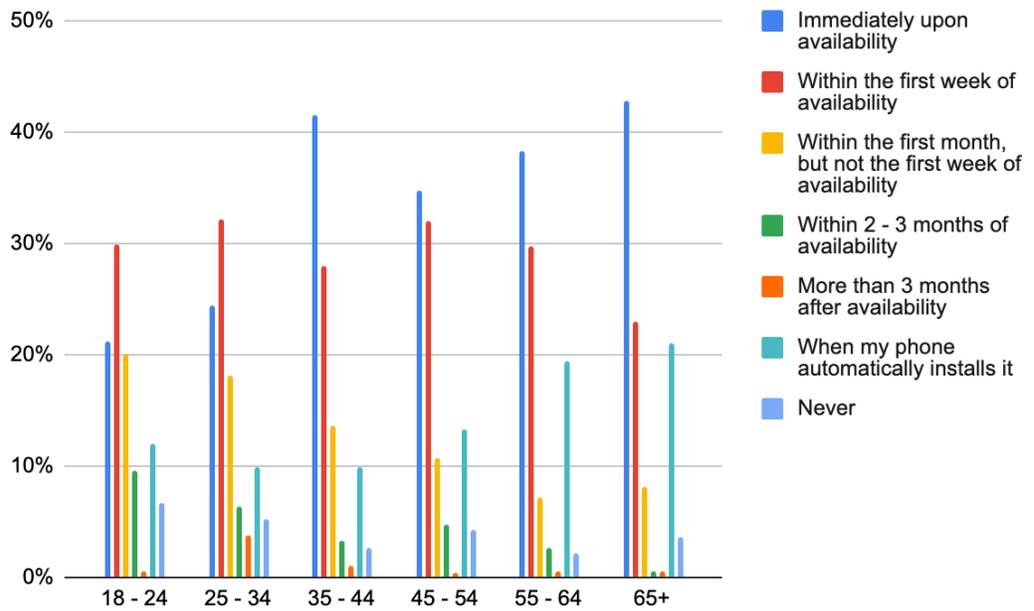

Figure 10: Security update frequency by age, United States (Figueroa, 2022)

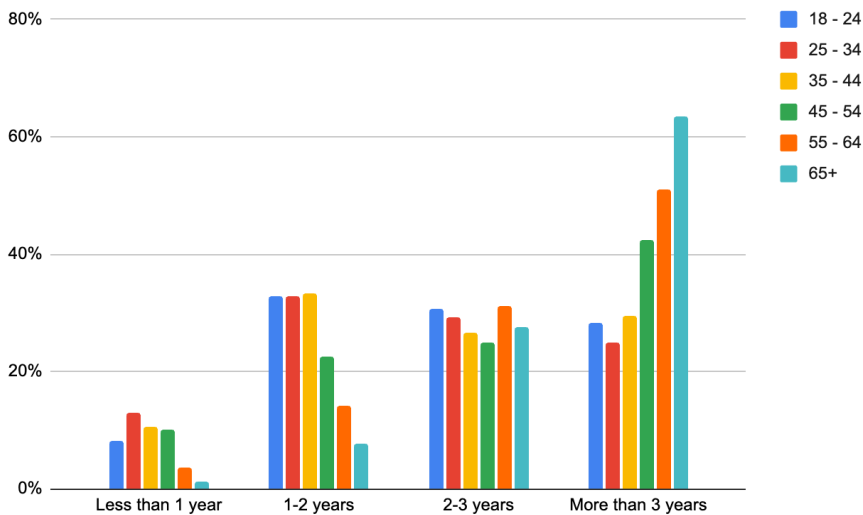

Figure 11: Length of smartphone Usage before replacing by age, United States (Figueroa, 2022)

3. A manual analysis will be required for apps that an ML analyzer needs help to classify. The quality of manual analysis highly depends on the experience and available time of the analyst. Misclassification may be introduced if the analyst needs to become more familiar with certain malware behaviors or the programming language used in the malware. So, bias may also be introduced by humans during this analysis stage.

Studies also confirmed that bias in ML models can cause significant performance degradation if the detection environment differs from the training environment. All of the biases above can impact malware detection efficacy along the lines of difference, yielding uneven or insufficient protection quality for users based on their demographics or other factors. However, there is limited research on biases from a product inclusion perspective. Many reports published by researchers and vendors show that malware infection rates vary significantly in different regions or user groups. Still, there is limited research on whether the different infection rates are connected with bias in the training data.

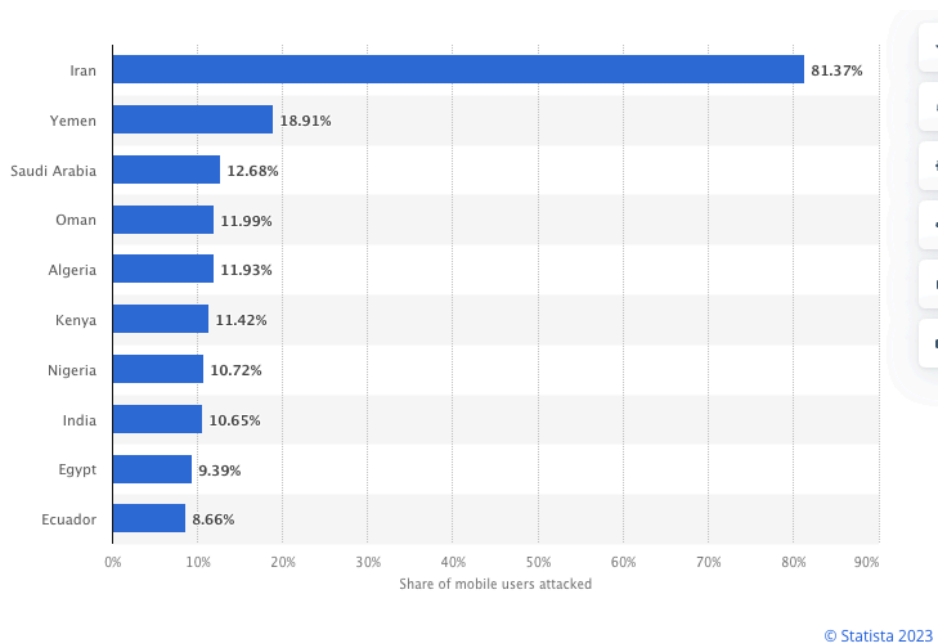

Figure 12: Countries with top malware Infection rate in Q3 2022 (Petrosyan, 2023)

It should be noted that many other factors can affect malware detection rates, such as varying security policies across app stores and users' security awareness. It is unclear whether ML fairness significantly affects the detection rate. However, given the importance of malware detection for users and the lack of research on the impact of bias on protection across user groups, we think that it is worth investigating this role and methods to reduce these biases to provide more robust and fairer protections against malware for all users.

2.3.5 Discussion

- As ML is increasingly used in security and safety systems, the ML capability must perform equally well for all users. Otherwise, the ML capability may underperform or make wrong decisions for users in certain groups, causing harm ranging from service inconvenience to financial and privacy losses due to cyber-attacks or even physical harm caused by downstream safety degradation in physical machines and infrastructure.
- Although many studies have been performed, not all topics are well explored. Some concepts, such as fairness for fingerprint systems, fairness in anti-spoofing detection, and whether bias plays a role in the malware infection rates in different regions and user groups, have been rarely examined.
- Many ML fairness challenges remain; some are generalized, such as the need for theoretical frameworks to measure fairness. Some are security- or safety-specific, such as a better understanding of the downstream safety impacts of false positives and negatives in ML systems.
- Other challenges include the need to study biases by demographic and non-demographic covariates, combinations of different covariants, and social impact from bias in ML-based security and safety systems.

# 3 Product Inclusion: Stalkerware

## 3.1 Intimate Partner Violence and Intimate Partner Surveillance

The American Center for Disease Control and Prevention (CDC) defines Intimate Partner Violence (IPV) as "abuse or aggression that occurs in a romantic relationship." Intimate Partner Surveillance (IPS) is a method in IPV where one partner monitors the location and behavior of the other through technical or non-technical means.

The 2016/2017 CDC report "The National Intimate Partner and Sexual Violence Survey" shows that, in the US, 13.5% of women and 5.2% of men have been stalked by an intimate partner. About 16 million women and 11 million men experienced sexual or physical violence or stalking from an intimate partner before age 18. Social and structural conditions contribute to inequities in risk for violence, with marginalized youth, such as sexual and gender minority youth, at a greater risk.

Technology can support IPS in various ways, including dedicated stalking software, dual-use software like Find My Phone apps, and without additional software. This section evaluates the current state of technology-enabled IPS on mobile phones and addresses challenges in developing products to protect users targeted by IPS.

The same mobile device-enabled IPS techniques can also be used in stalking where the perpetrator is not in an intimate relationship with the victim. Still, we believe that stalkers typically use stalkerware to target people who are physically and emotionally close to them.

## 3.2 Stalkerware as a means to Intimate Partner Surveillance

Stalkerware, or commercial spyware or spouseware, is software specifically intended to track and control another person's mobile activity. It's commonly used for monitoring or controlling purposes by someone who knows the target personally. Unlike spyware, stalkerware doesn't collect data indiscriminately from many devices but is targeted to monitor a specific individual.

Internet searches for phrases like "spy on wife software" or "catch cheating spouse phone" reveal a thriving ecosystem of stalkerware for all modern operating systems. Stalkerware enables users to monitor various phone features such as reading SMS or messaging apps like WhatsApp, tracking location, accessing photos, and listening in on calls. These capabilities are often exploited to control or manipulate the targeted user.

## 3.3 Google's restrictions on stalkerware

Google takes stalkerware seriously and restricts its services and software against its usage. For example, since 2020, Google's advertising products may not be used to advertise stalkerware. The Google Play store also bans stalkerware apps.

The Android Security team defines stalkerware apps as potentially harmful applications (PHA) in [Android's published PHA categories](#) as follows:

> *Code that collects and/or transmits personal or sensitive user data from a device without adequate notice or consent and doesn't display a persistent notification that this is happening.*

> *Stalkerware apps target device users by monitoring personal or sensitive user data and transmitting or making this data accessible to third parties.*
>
> *Apps exclusively designed and marketed for parents to track their children or enterprise management, provided they fully comply with the requirements described below, are the only acceptable surveillance apps. These apps cannot track anyone else (a spouse, for example) even with their knowledge and permission, regardless of whether persistent notification is displayed.*

The Android Security Team classifies stalkerware as potentially harmful, removes it from Google Play, disables it on devices, and warns users who installed it.

## 3.4 Stalkerware as an exceptional malware category

Stalkerware poses unique physical and psychological dangers to targeted users. Based on our research and our interviews with domestic abuse support groups, its presence is often known to the target, but fear of retaliation prevents removal. Abusers often even disclose the app's presence as part of the abuse cycle, and attempts to remove it can result in physical harm or even death. Other targeted people may sense that they are under surveillance but don't have the technical knowledge to understand how they are monitored.

While Google Play Protect warns targeted users and turns off stalkerware, preventing escalation to physical violence distinguishes stalkerware from all other malware categories and is crucial in our effort to protect users from stalkerware. This paper explores additional ways to help users achieve safety on their phones and personal lives.

## 3.5 The challenge of dual-use software

If Google detects and prohibits stalkerware on Google Play and Android phones, why isn't this the end of the story? Despite Google's efforts to remove stalkerware from Google Play and Android phones, other legitimate apps with similar capabilities exist. These include parental control and corporate apps for tracking vehicles, employees, or stolen phones.

These apps provide similar functionalities to stalkerware, often allowing the stalker to read SMS, track location, and take photos. The critical difference lies in consent between the tracker and the tracked or a legal framework permitting tracking under specific circumstances.

The debate around these dual-use apps centers on their permitted functionality and availability in app stores, with many people having different opinions on this topic. The Android Security Team needs to strike a balance between permissible and prohibited uses. Still, regardless of where mobile platform policies draw this line, not all stakeholders will agree on where that line is drawn.

## 3.6 Prior stalkerware research

Academics, security companies, and independent experts extensively research stalkerware. The tech press and popular press also warn users regularly about these applications and their potential for abuse. They also warn users about how stalkerware is unique compared to other malware due to the personal connection between stalker and victim, elevating it to a higher level of sensitivity.

The [Coalition Against Stalkerware](), formed in November 2019, aims to combat stalkerware by combining the efforts of security companies, domestic abuse support groups, digital rights advocates, and law enforcement agencies. Over the years, it has become a valuable resource for stalkerware-related information and research.

In April 2022, Kaspersky, a security company in the Coalition against Stalkerware, reported a 39% drop in stalkerware installations on phones with their anti-malware software. Yet the report cautions that this decline is only a fraction of the problem, as over a million Android devices may be under surveillance.

Cornell and NYU researchers aided stalkerware victims and reported their findings in a paper titled "The Many Kinds of Creepware Used for Interpersonal Attacks". Using novel techniques to discover stalkerware and dual-use apps, they found over 1,000 malicious apps and reported them to the Google Play team. In another study, "The Spyware Used in Intimate Partner Violence," they found that 20% of domestic abuse victims had their phones monitored.

## 3.7 Stalkerware on Android

Android, the world's most popular operating system, plays a significant role in the lives of billions of people. Regrettably yet nearly inevitably, stalkerware exists on Android, affecting approximately 1.5 million annually, especially in the USA, Russia, and France. This holds even after adjusting for population.

Stalkerware has been banned from Google Play since 2008 under the generic anti-malware policy. In 2016, Google Play published detailed malware definitions that banned stalkerware using the formerly used term commercial spyware. In 2020, the malware policy was again updated to rename commercial spyware to stalkerware.

Google Play allows dual-use apps with stalkerware-like functionality, but the dual-use apps policy mandates several requirements to reduce the risk of abuse:

- Apps must not present themselves as a spying or secret surveillance solution.
- Apps must not hide or cloak tracking behavior or attempt to mislead users about such functionality.
- Apps must always present users with a persistent notification when the app is running and a unique icon that clearly identifies the app.
- Apps and app listings on Google Play must not provide any means to activate or access functionality that violates these terms, such as linking to a non-compliant APK hosted outside Google Play.
- You [the developer] are solely responsible for determining the legality of your app in its targeted locale. Apps determined to be unlawful in locations where they are published will be removed.

The Android Security team has fortified the OS against stalkerware, making it harder for such apps to operate. Other changes increased user awareness of dual-use apps.

The Device Administration API [was deprecated in Android 9](), preventing stalkerware from gaining privileged device access, for example, to stop targeted users from uninstalling the app. Replacement APIs like Work Profile Mode or Device Owner Mode put users in charge of these privileges, protecting them from stalkerware abuse.

As stalkerware requires many permissions, we improved auditing capabilities to help users better understand app permissions. We added a visible indicator when an app uses the camera or microphone. Tapping the indicator shows which app is using the permission. The Privacy Dashboard, introduced in Android 12, provides a consolidated UI to see which apps recently used permissions. This allows users to audit which apps recently used a microphone, camera, or other sensitive permissions. For users who would instead disable the microphone or camera completely, [Android 12 also introduced quick toggle buttons to do so](#).

Android 11 introduced auto-resetting permissions for unused apps to reduce the burden on users. After a few months without user interaction, the OS resets granted app permissions, making stalkerware useless even if it's permanently running in the background. Android 12 improved this feature by putting unused apps into hibernation, revoking their permissions, and limiting their background activities.

Android 8 [introduced restrictions on background apps accessing GPS locations](#). Android 10 gave users more control over location access by introducing the ACCESS_BACKGROUND_LOCATION permission. This puts the user in charge of granting this permission and makes it harder for stalkerware to track users' locations.

Many stalkerware apps tried to hide their existence by hiding launcher icons, but Android 10 introduced changes that made this impossible.

## 3.8 Additional Product Inclusion Challenges with Stalkerware

Android and Google Play continue to evolve to better protect users from stalking through dual-use apps or stalkerware. Key questions to be answered are (1) where to draw the line between prohibited stalkerware and permitted dual-use, (2) what to ask of legitimate dual-use app developers to minimize the potential for abuse, and (3) how to improve the protection of targeted users.

### 3.8.1 Google Play

The line between stalkerware and dual-use apps on Google Play is evolving, but as long as dual-use apps are permitted, our product inclusion challenges and solutions must work with those apps in mind.

Exploring further measures to curb abuse by legitimate dual-use app developers may be effective. As noted before, the Google Play policy already mandates many requirements. With creativity, apps can improve further. One example of a dual-use app that does more than Google Play policy requires is Google Maps. Millions of users use its location-sharing feature to share real-time location information with family members or friends. Still, it can also be abused by stalkers to track their targeted users. To mitigate this, Google sends monthly reminder emails to all parties involved in location sharing. This can alert the targeted user if they are unaware that the feature is enabled.

### 3.8.2 Options for the Android Operating System

Asking app developers to be honest about the abuse potential of their dual-use apps relies on the collaboration of these developers. Adjusting the Android operating system to safeguard users from dual-use apps may be more efficient.

#### Options for sideloading

After Google Play's initial crackdown on stalkerware over a decade ago, the thriving stalkerware ecosystem migrated off Google Play. Now, stalkerware app downloads and installations occur through individual app websites or third-party stores that allow them.

Sideloading is the process of downloading and installing apps from sources other than Google Play. Sideloading initially requires user confirmation to proceed due to the higher risk of malware. To enable it, users must adjust settings. Google Play Protect scans sideloaded apps to warn users about potential threats.

Notably, the defenses against sideloaded malware do not effectively guard against stalkerware. In almost all instances, the stalker physically accesses the user's device to install the stalkerware. Before installation, the stalker can easily activate sideloading settings or turn off Google Play Protect's malware defenses.

Improving sideloading should account for the stalker's likely physical access to the user's device. An existing option is Google's Advanced Protection program, offering enhanced security by turning off sideloading, except through ADB, which demands higher technical expertise. Disabling Advanced Protection requires the user's Google account password, triggering notifications that could alert the targeted user to tampering attempts.

### Options for installed stalkerware

Suppose stalkerware is already installed on a targeted user's device. In that case, available protection options fall into two categories: it may be possible to limit the stalking, or it may be possible to improve user awareness.

Notifying users of potential stalkerware on their devices is challenging due to the stalkerware's presumed control. Traditional security features and notifications can be turned off or dismissed by the stalkerware or its operator. An alternative, out-of-band solution is needed to address this.

Discreet notifications about stalkerware could be shown on connected devices like smartwatches or computers. This would hinder manipulation by stalkerware, even for stalkers with physical device access.

Another out-of-band option is surveillance warning banners in Google apps like Gmail. Due to Android's app isolation guarantees, installed stalkerware apps could not manipulate or suppress these warnings. Google already uses this technology to warn users in Gmail if nation-states are attacking their accounts.

Another option may be to limit the abuse risk of dual-use apps. As these apps are still distributed through Google Play, Google could introduce additional requirements that are unavailable for sideloaded apps. For example, as of 2023, all Google Play apps that may be abused for stalking are asked to self-declare as monitoring applications through [the "IsMonitoringTool" flag](). The operating system could warn users of installed apps that hold this flag in the future.

Limiting stalkerware functionality through technical means is more difficult as the stalkerware (or its human operator) may turn off any relevant security or privacy features that seek to limit its functionality.

Even if security protections are enabled, users may still not notice warnings. To reduce the risk of stalkerware, better ways to notify users should be explored, such as when they're actively using their phones. This can help prevent notifications from being missed in the overflowing notification bar.

## 3.9 Helping targeted users reach a safe state

The challenges discussed so far are technical, but more than technical solutions are needed to solve non-technical problems. The ultimate goal of any stalkerware protection should be to help users achieve physical and psychological safety from their stalker.

One potential solution is a warning that informs users that their phone is compromised, cannot be trusted, and they should seek help. Targeted users need to realize that their phone is compromised and cannot be trusted anymore and that they should seek assistance from trusted friends, support groups, or law enforcement without using their phones.

The previously mentioned "IsMonitoringTool" flag provides visibility to users with dual-use apps downloaded on their Android phones. A notification on the phone's safety center informing the users of the app could be present as long as the app remains downloaded and active on the phone.

We are also developing platform capabilities to monitor installed apps for invasive patterns. If an app is found to be potentially abusive, we report it to Google and warn users via Play Protect.

The crucial point is not necessarily to uninstall the stalkerware app, as doing so might escalate harm to the victim, as explained earlier. Instead, the focus is on notifying the user of the surveillance and suggesting next steps. Once the targeted user is aware of the surveillance, they can try to find victim support groups with the help of trusted friends or relatives.

## 3.10 Conclusion

Most of the stalkerware product inclusion challenges on Android come from the existence of dual-use apps and the physical proximity of the stalker and targeted user. Inside and outside Google, future product strategy must consider these conditions when seeking to protect targeted users from Android-based stalking. A two-pronged approach with different measures may be helpful to protect from dedicated stalkerware and protecting from repurposed dual-use apps.

# 4 Product inclusion challenges in cellular security

All mobile devices that connect to a cellular network are vulnerable to false Base Station (FBS) attacks. There is no cryptographic way for a [smartphone to verify either the legitimacy of a base station or the integrity of the](#) messages it transmits. This applies to [2G](#), [3G](#), [LTE](#), and [5G](#), and it is a systemic cellular protocol issue, not an OEM or carrier issue. Moreover, this systemic risk cannot be mitigated or addressed by, for example, E2EE and/or a VPN.

Although all mobile users are vulnerable to FBS-based threats, these attacks are known for disproportionately impacting at-risk users.

FBS has been abused in the military conflict in Ukraine for the [distribution of propaganda](#), [electronic warfare](#), and [surveillance of Ukranian militias](#). Stingrays ([FBS-based surveillance tools](#)) have been widely reported in the context of [dragnet surveillance against at-risk users](#), including documented cases in the US of tracking and surveillance of [undocumented immigrants](#) and attendees to Black Lives Matter protests. These racial disparities in police 'stingray surveillance' [have been analyzed in recent years](#). Most recently, leaked documents provided insight into how the Iranian government systematically abuses these techniques to [track and control the phones of protesters](#).

FBSs were also identified by Amnesty International as [one of the main techniques leveraged to deploy the Pegasus malware](#) onto the phones of political dissidents, journalists, and activists.

The majority of at-risk populations, most likely to be the subject of FBS attacks, are users of low-cost devices. These most vulnerable individuals, with clear indications of being actively targeted by these attacks, regularly use old and low-cost devices that are likely to run older, unpatched versions of the OS. Meanwhile, the majority of premium users both use much more secure devices and are way less likely to be the target of a FBS attack.

Besides FBS attacks, cellular users are also exposed to other attack vectors. For example, [recent studies](#) revealed that some carriers do not encrypt cellular traffic over the air, actively exposing circuit-switched voice calls and text messages to dragnet surveillance. There is also strong evidence that a particular cipher of a legacy cellular protocol [was almost certainly backdoored intentionally](). Even though legacy protocols and circuit-switched voice and messages are rarely used, they are heavily used and, in some cases, the only available connectivity option in many underdeveloped nations. This further deepens the disproportionate risk of cellular attacks for at-risk users.

To make matters worse, this disproportionate imbalance in the cellular security threat model may disincentivize the telco industry to prioritize FBS attack risk. Although it is not necessarily correlated, FBS risk has remained mostly unaddressed throughout all mobile generations. It is also very challenging to motivate work to harden the firmware running in the cellular baseband when most vulnerable users have older devices with operating systems that no longer receive security updates.

# 5 Conclusions and Future Work

This survey seeks to raise awareness of the challenges and need for more research on product inclusion in digital security and privacy. As discussed at the beginning of this work, product inclusion and product security share a similar challenge and approach to improvement: we must think expansively about how products may fail to offer an inclusive or secure experience and respond with new objectives to fill the discovered gaps. Doing this threat modeling proactively and as part of the product development life cycle, rather than reactively due to production failures, is a well-established best practice to reduce harm and improve user experience. While this concept of threat modeling is common in security, it is less common in product inclusion. In particular, this paper focused on the need to apply threat modeling more effectively in product inclusion, which has not received expansive thought and research: inclusive security and privacy.

This paper is not, on its own, an attempt to cover product inclusion for security and privacy in that expansive sense. We introduced a broad set of potential areas of concern and some interesting consumer survey results relating to security and privacy across countries, observing inclusion differences driven by economic and cultural differences. We then explored a deeper dive across several areas: ML fairness, stalkerware, and cellular security.

Some areas where we think further research should be prioritized: measuring the loss of security and privacy caused by inclusion gaps; further exploration of impact across additional demographic dimensions, such as children and elderly (beyond the biometric example cited in this paper); effects on mobile user education levels; impacts on at-risk populations (beyond the stalkerware and cellular security examples in this paper); impacts on first-time Internet-connected users; impacts related to mobile phone-adjacent digital technologies such as tracking tags and their associated networks; additional exploration of device theft risks; impacts of the preloaded software content on mobile devices ; impact and prevalence of mobile device-enabled social engineering attacks and defenses; impact in variations of malware techniques employed; and inclusive security and privacy in other areas of digital technology not specific to mobile devices (e.g., IoT sensors, networking, Internet use and digital literacy, impact of digital security and privacy regulations, etc.).

Again, the above suggestions for future research are not exhaustive. Still, we hope this survey will motivate further research and work in this critical and historically underserved area of responsible digital technology development.

## Acknowledgments

We would like to extend our sincere thanks to Khawaja Shams, Eugene Liderman, Salvador Mandujano,  Mina Askari,  William Luh, Elie Bursztein, Sunny Consolvo, Matthew Murray, Manya Sleeper, and Olivier Tuchon, who generously offered their time and knowledge to support this research. Your help significantly improved this research.